# Spatial organisation of French research from the scholarly publication standpoint (1999-2017): Long-standing dynamics and policy-induced disorder

*Michel* Grossetti[2], *Marion* Maisonobe[1,*], *Laurent* Jégou[2], *Béatrice* Milard[2], et *Guillaume* Cabanac[3]

[1]CNRS, UMR Géographie-cités, 93300 Paris - Aubervilliers, France
[2]Université de Toulouse, UMR LISST, 31058 Toulouse, France
[3]Université de Toulouse, UMR IRIT, 31062 Toulouse, France

**Abstract.** In social processes, long-term trends can be influenced or disrupted by various factors, including public policy. When public policies depend on a misrepresentation of trends in the areas they are aimed at, they become random and disruptive, which can be interpreted as a source of disorder. Here we consider policies on the spatial organization of the French Higher Education and Research system, which reflects the authorities' hypothesis that scientific excellence is the prerogative of a few large urban agglomerations. By geographically identifying all the French publications listed in the Web of Science databases between 1999 and 2017, we highlight a spatial deconcentration trend, which has slowed down in recent years due to a freezed growth of the teaching force. This deconcentration continues, however, to sustain the growth of scientific production in small and medium-sized towns. An examination of the large conurbations shows the relative decline of sites that nevertheless have been highlighted as examples to be followed by the Excellence policies (Strasbourg among others). The number of students and faculty has grown less there, and it is a plaussible explanation for the relative decline in scientific production. We show that the publication output of a given site depends directly and strongly on the number of researchers hosted there. Based on precise data at the French level, our results confirm what is already known at world scale. In conclusion, we question the amount of disorder resulting from policies aligned with poorly assessed trends.

## 1 Introduction

In France, the geographical organization of Higher Education and Research (HER) has long been marked, as with other activities, by a concern for territorial rebalancing. Since the decentralizations initiated by the Mendès-France government in the 1950s, this concern has been pursued by Gaullist governments, then resumed in the 1980s and the early 1990s.

However, starting in the 1990s, economic and geographic research groups, inspired by a more general discourse on "global cities" [1,2] and the New Economic Geography (NEG) [3–7], helped to recast this concern for territorial balance as a discourse on the differentiated competitiveness of territories and the need to strengthen France's only "global city": the Paris conurbation [8–13].

As far as HER is concerned, this change of perspective cooccurred with the launch of the Shanghai ranking in 2003 by a team from Shanghai's Jiao Tong University. This university ranking is mainly based on the number of Nobel Prize winners, publications listed in the Science Citation Index of Clarivate Analytics (formerly Thomson Reuters) and articles published in the *Nature* and *Science* journals.

Developed to benchmark Chinese universities against those of other countries and using the most prestigious American universities as a model, it logically ranks universities such as Harvard or Princeton top. The French universities specialised in the natural sciences and technology rank way lower.

Although heavily criticised [14], this ranking has inspired the so-called "Excellence policies" aimed at avoiding "spreading the means too thinly"[a], affirming (in a brief) that research "is not intended to support regional planning, but must aim for excellence". Quite surprisingly, the perspective of the reformers at the end of the nineteenth century who wanted to establish a small number of large universities that could compete with their German counterparts has been revived a century later [15].

---

[a] "*I see real demons coming back, the ones of sprinkling. But don't let this notion of excellence in higher education be killed!*" (Laurent Wauquiez, Minister of Higher Education, at the announcement of the results of the second round of selection of "Excellence laboratories", 14 February 2012).

---

[*] Corresponding author: marion.maisonobe@cnrs.fr.

Launched by the French government in 2010, the "Initiatives d'excellence" or Idex[b] and their variations ("Laboratoires d'excellence" a.k.a. Labex, "Équipements d'excellence" a.k.a. Équipex, etc.) were funded by a part of the "great loan" (7.7 billion euros). Most of the accepted projects concentrated in the Paris region (4 Idex out of 8 and 84 Labex out of 171) and a few major peripheral scientific centres (Bordeaux, Toulouse, Aix-Marseille, Strasbourg benefited from Idex from the first round). Other sites subsequently benefited from Idex and I-sites (a similar but less well-endowed system), with the city of Toulouse losing its own after the failed attempt at merging the 3 universities of Toulouse. In parallel with these policies allocating financial resources to a few pre-selected sites only, the recruitment of HER civil servants [16] slowed down in recent years. As a result, more temporary positions were offered to meet the needs in terms of research, and more particularly, in terms of teaching as the student population grew significantly [17–19].

As reported earlier [20], the spatial logics of these policies are based on a series of beliefs that are not specific to France and which concern in particular geographical concentration and its supposed effects on the "productivity" of researchers. Concisely, co-locating researchers (i.e., increasing their geographical concentration) is supposed to increase their "productivity". It follows that the best researchers and the best students head to large sites, thus reinforcing the geographical concentration of research. This phenomenon known as "critical mass" or "agglomeration effect" would therefore strengthen the hierarchies between major first-class sites, capable of being integrated into the most efficient international networks, and the others, condemned to second-class research or simple assistance to networks run from the major centres. This distinction would legitimize the French political plan to enforce a growing differentiation between research sites with a global vocation and smaller sites operating higher education and local/regional development only [21,22].

However, our team has been studying the global geography of research activities for about twenty years, notably through publications indexed in the Web of Science, and our results do not support the thesis of a spontaneous concentration of higher-level research in metropolitan areas. In particular, our previous work has shown a global trend towards deconcentration between cities, at the global and country levels, of publications, collaborations, and citations obtained through publications [23–25]. These results further suggest that "scientific output", however measured, for a given geographical delineation (cities, regions, countries), is a function of the number of researchers and does not depend on any city size effect.

What has been the situation in France in recent years? Any geographical re-concentration resulting from Excellence policies? Any effect of concentration, and additional resources allocated to major sites, on the "productivity" of researchers? To verify this, we studied France's research output using scientific production data from 1978 to 2017 and available data on the number of researchers per urban agglomeration.

In the following sections, we stress the contradictions between long-term trends and recent policies by focusing on two points. First, the trend towards the geographical deconcentration of research activity contradicts policies that take for granted the existence of an inverse trend that they seek to accompany or reinforce. Second, the lack of effect of the geographical concentration of researchers in the largest agglomerations on the scientific productivity of these agglomerations contrasts with policies hypothesising this effect. We conclude by contrasting the regularity of observed trends with the disorder introduced by policies based on their misinterpretations.

## 2 Changes in the "French scientific map": a continuous deconcentration

Data from large bibliographic databases offer a global vision of the territorial distribution of scientific activities. We have therefore coded the geographical location of affiliations declared by the authors of publications listed in the Science Citation Index Expanded (SCI), the reference index of the Web of Science. This index is restricted to the natural and technical sciences, which is the scope of the analysis presented in this section. The geographical coding of scientific information was performed with higher precision than in most current similar studies.

First, we grouped the authors' cities at the level of urban agglomerations, which avoids the effects of deformation due to variations in municipal boundaries and enables the representation of geographically homogeneous entities [26]. Second, we accounted for the publications by fractionating the values attributed to the different urban agglomerations. Each publication represents a unit (one point) divided uniformly between the different urban agglomerations present in the byline. For example, a publication co-authored by authors from the Paris conurbation and others from the Nantes conurbation will translate to 0.5 point to the Paris conurbation and 0.5 point to the Nantes conurbation[c]. This way of counting avoids overestimating the activity of scientific centres that collaborate a lot with others while normalising disciplinary-wise practices in terms of co-signing articles. In the following tables, we join the recent data available (1999-2017) with data processed earlier (1978-1988) in order to track decade-long changes and thus better identify long-term trends. While some of the underlying geographical boundaries have been slightly adjusted between the 78-88 series and the 99-2017 series, the process chain used (geocoding, spatial aggregation, fractional counting at urban agglomeration level) is identical.

---

[b] "*The "Initiatives of Excellence" (Idex) program aims, by making world-class research a lever and a driving force, to create 5 to 10 multidisciplinary centres of excellence in higher education and world-class research in France. These clusters will be organized in the form of territorially coherent groups of higher education establishments, universities and schools, involving research organizations and in partnership with businesses. They will be built on scientific forces of excellence, multidisciplinary and internationally recognized, and on innovative research and training activities, all developing around attractive campuses of the highest international standards.*" Excerpt of the call for tenders "Initiatives of excellence", 2010

[c] Denis Eckert, Laurent Jégou and Marion Maisonobe carried out this work in collaboration with Yves Gingras and Vincent Larivière of the CIRST unit in Montreal.

## 2.1. Production and workforce throughout the territory and time

Table 1 shows the evolution of the spatial distribution of French publications obtained by the aforemnetioned method. We grouped the agglomerations into four types to cross the French administrative hierarchy (capitals of regions or similar administrative entities called *départements*), which partially covers the hierarchy of size, the hierarchy of higher education institutions (presence or not of a full-fledged university) and the effects of inertia of the historical process of construction of the French university system: the historical hierarchy — distinguishing the academic centres of 1870, to which we included Strasbourg, which had faculties before the War of 1870 and a German university between 1870 and 1914.

**Table 1.** Share of scientific output by type of agglomeration (SCI Exp, split by agglomeration)

| Year (3-year moving average) Type of urban areas | 1980 % | 1988 % | 2000 % | 2008 % | 2016 % |
|---|---|---|---|---|---|
| Paris | 46.1 | 44.7 | 36.8 | 34.1 | 32.9 |
| Other academic centres 1870 | 41.2 | 41.7 | 44.8 | 45.5 | 45.4 |
| Other university sites | 10.4 | 11.6 | 14.4 | 15.7 | 16.3 |
| Other cities | 2.3 | 2.1 | 4.0 | 4.7 | 5.3 |
| Total | 100.0 | 100.0 | 100.0 | 100.0 | 100.0 |

Source: 1979-89 (OST-Montréal-LISST, 2012), 1999-17 (IRIT-LISST, 2019). Metropolitan France, Overseas Departments and Territories.

Table 1 highlights the decline of Parisian hegemony, which accelerated between 1988 and 2000, under the effects of the massification of the student population that marked this period. This decline results from greater recruitment of faculty members and researchers outside Paris, in the universities, but also at the CNRS (i.e., the French national research center) and other national establishments, as shown in Table 2. In particular, Greater Paris' share of student numbers under the responsibility of the Ministry of National Education fell from around 30.0% in 1993 to 26.4% in 2012. The proportion of tenured faculty (university professors, lecturers, and secondary school teachers working in higher education) fell from 25.8% in 1992 to 24.3% in 2012 [19,27,28].

For the CNRS: "*The distribution of the State subsidy in 1990 shows that the Île-de-France region [Greater Paris] still receives just over half of the resources. However, the best progress has benefited the other regions, particularly Limousin (+16% annual growth between 1980 and 1990), Pays de la Loire (+10.4%), Picardie (+10.1%) and Nord Pas de Calais (+8.7%). However, the best-endowed regions remain Rhône-Alpes, PACA and Alsace. Staffing reproduces the same balances. Île de France hosts 52% of the total workforce, followed far behind by Rhône-Alpes (9.4%), PACA (8.3%) and Alsace (5.9%). The Western regions remains at a disadvantage. The rate of researchers in the provinces reached the 50% mark in 1993, and rose to 53 per cent in 1996.*" (pp. 395-396) [29].

However, a more recent census of the CNRS workforce, conducted in December 2019, indicates that the Paris region (Île-de-France) still accounts for 39.8% of researchers, only 0.2 points less than in 2010-2013, suggesting a slowdown in the process of spatial deconcentration between these two dates. In particular, there is a slight re-concentration in the Paris region for the chemistry and the human and social sciences institutes.

**Table 2.** Change in Paris region's share of permanent staff in various research organisations

| Research organisations | Share of Paris region (1988-1990) in percent | Share of Paris region (2010-2013) in percent |
|---|---|---|
| CNRS | 53 | 40 |
| INRA | 30 | 26 |
| INSERM | 60 | 52 |
| CNES | 19 | 19 |
| CIRAD | 45 | 40 |
| INRIA | 62 | 35 |

Source: « Bilans sociaux » 2010 et 2011, Grossetti, 1995 et Mailfert, 1991.

Table 2 shows the stagnation in staffing levels in most public research organisation from the 1980s onwards. For example, the CNRS, which had just over 24,000 members in 1986, was around 22,000 in 2013, after a few fluctuations. The CEA had a little over 14,000 members in 1989, and its workforce exceeded 16,000 in 2011. The INRA had almost the same number of staff in 2013 as in 1989 (8,376 then and 8,576 now), as does the CNES (from 2,310 to 2,330). IFREMER has grown from around 1,200 staff in 1988 to 1,300 in 2009. CIRAD has lost staff (from 2,300 in 1990 to just under 1,800 at present) and IRSTEA, which replaced CEMAGREF and has just merged with INRA, has slightly fewer permanent staff (952 in 1989, 872 at present).

Only INSERM (5,200 in 2010 compared to 3,700 in 1989) and INRIA (1,347 in 2012 compared to 592 in 1989) have seen a significant increase in staff numbers[d].

---

[d] These figures taken from the social balance sheets or AERES assessments concern permanent staff, calculated in "full-time equivalent" where the data allow, which is not always the case.

All of these organizations now have high levels of non-permanent staff (doctoral students, post-doctoral students, fixed-term contracts, various temporary staff), for which any comparison with the situation of 20 or 30 years ago is difficult because non-permanent staff were generally not accounted for at that time. However, it can be hypothesised that non-permanent employment has increased in most organizations, as it has increased in higher education [18], which corresponds well to the reports of growing precariousness in the HER professions which has been expressed several times over the last twenty years [17].

This relative stagnation of the national research organisations contrasts with the increase in the number of university staff, which followed the increase in student numbers, albeit with a delay and to a lesser extent, but which was significant, at least until 1998. However, here too, particularly in the human and social sciences domains, recourse to precarious staff has become massive [30]. Similarly, Cytermann et al. estimated a 40% increase in student numbers between 1992 and 2002. They also note a quicker increase in numbers in small university towns, universities on the outskirts of Paris and universities of litterature, humanities, and social sciences, and less in universities in downtown Paris and those in large cities that specialise in the material and technical sciences [31].

In other words, since the early 1970s, all academic research institutions have tended to deconcentrate at the national scale, but in the case of academics, this spatial deconcentration occurred with an increase in enrolments, and thus an institutional rebalancing in their favour.

## 2.2. Evolving scientific output by type of site

Table 1 shows that the decline of the concentration of publications in the Paris region corresponds to the growth in the weight of the major old academic centres, other university sites, which are often regional or departmental capitals with universities, as well as smaller sites, which fall into the category of "other cities". A word of caution here: researchers hosted at establishments in the latter category tend to sign their publications using the address of their home university, located in another (bigger) city. It should also be noted that some universities are organised in a network or alliance[e] and that assigning the university to a single city, as we do here, can be a somewhat misleading approximation.

Now let us look at the oldest university cities (Table 3). We separated the 10 most important ones by number of students, researchers, and publications from

---

We have often rounded them off when different documents present different counts. They should be considered as approximations, although being sufficient to convey the general idea of trends.
[e] For example, the universities of the Littoral (Boulogne, Calais, Dunkirk, Saint-Omer), of Artois (Arras, Béthune, Douai, Lens, and Liévin), or of Savoie (Annecy, Chambéry, Le Bourget-du-Lac, and Jacob-Bellecombette).

---

the 5 that are smaller in size. Among the top 10, growth is almost general with the very notable exception of Strasbourg, whose contribution to French publications is declining (from 4.6% to 2.6% of national production between 1980 and 2016), and Lyon, Marseille, and Nancy, which are stagnating.

Strasbourg is an intriguing case and we have no definite explanation. We simply note that the relative decrease in publication output from this city applies to selected sectors of biology. The proportion of publications attributed to Strasbourg is close to the national figure for cities hosting that many students.

Among the five medium-sized centres, Clermont-Ferrand and Besançon are the only ones whose contribution to French publications is increasing.

It should be noted that some of these agglomerations have lost their regional monopoly relatively recently: this is the case, for example, of Poitiers since the creation of the University of La Rochelle in 1993.

**Table 3.** Shares of scientific production of agglomerations of the type "academic centres in 1870" (SCI Exp, fractioned by agglomeration)

| Year (3-year moving average) Agglomeration | 1980 % | 1988 % | 2000 % | 2008 % | 2016 % |
|---|---|---|---|---|---|
| *Major centres* | | | | | |
| LYON | 6.57 | 5.78 | 5.89 | 5.88 | 6.01 |
| GRENOBLE | 4.51 | 4.86 | 5.66 | 6.27 | 5.62 |
| TOULOUSE | 3.78 | 4.03 | 5.20 | 5.48 | 5.51 |
| MARSEILLE | 4.21 | 3.92 | 4.02 | 4.16 | 4.28 |
| MONTPELLIER | 3.29 | 3.61 | 3.85 | 3.97 | 3.89 |
| BORDEAUX | 2.81 | 3.10 | 3.25 | 3.27 | 3.47 |
| STRASBOURG | 4.56 | 3.99 | 3.50 | 2.75 | 2.62 |
| LILLE | 2.60 | 2.63 | 3.07 | 3.15 | 3.34 |
| RENNES | 1.56 | 1.78 | 2.33 | 2.82 | 2.97 |
| NANCY | 2.62 | 2.53 | 2.45 | 2.26 | 2.21 |
| *Medium-sized centres* | | | | | |
| CLERMONT-FERRAND | 1.21 | 1.43 | 1.53 | 1.43 | 1.46 |
| DIJON | 1.17 | 1.16 | 1.18 | 1.15 | 1.19 |
| CAEN | 0.66 | 0.96 | 1.10 | 1.09 | 1.00 |
| POITIERS | 0.89 | 0.97 | 0.97 | 0.92 | 0.88 |
| BESANCON | 0.61 | 0.77 | 0.83 | 0.89 | 0.99 |

Source: 1979-89 (OST-Montréal-LISST, 2012), 1999-17 (IRIT-LISST, 2019)

If we now look at the other regional capitals, which were endowed with a university in the 1960s (Table 4), it appears that Nantes has reached a rank in terms of scientific production that is now in line with its place in the urban hierarchy. Amiens has benefited from the general tendency to study longer as a student in recent years in the northern regions of France; Limoges has seen a slight increase; but the other conurbations are stable in terms of their share of scientific production.

**Table 4.** Shares of other regional capitals in scientific production (SCI Exp, fractioned by agglomeration)

| Year (3-year moving average) Agglomeration | 1980 % | 1988 % | 2000 % | 2008 % | 2016 % |
|---|---|---|---|---|---|
| NANTES | 1.04 | 1.21 | 1.87 | 2.02 | 2.28 |
| ROUEN | 0.96 | 0.90 | 1.04 | 0.77 | 0.90 |
| ORLEANS | 0.85 | 0.86 | 0.88 | 0.78 | 0.82 |
| REIMS | 0.60 | 0.62 | 0.68 | 0.61 | 0.65 |
| AMIENS | 0.38 | 0.41 | 0.47 | 0.58 | 0.59 |
| LIMOGES | 0.57 | 0.58 | 0.64 | 0.61 | 0.66 |
| AJACCIO[f] | 0.00 | 0.01 | 0.02 | 0.03 | 0.02 |

Source: 1979-89 (OST-Montréal-LISST, 2012), 1999-17 (IRIT-LISST, 2019)

Finally, let us look at departmental capitals with a full-fledged university (Table 5). Here again, similarly as for Nantes as seen above, large cities that were only equipped with universities in the 1960s and 1970s (Nice and Angers) have reached a level of scientific activity more in line with their place in the urban hierarchy.

**Table 5.** Shares of other departmental capitals with full-fledged universities in scientific production (*SCI Exp*, fractioned by agglomeration)

| Year (3-year moving average) Agglomeration | 1980 % | 1988 % | 2000 % | 2008 % | 2016 % |
|---|---|---|---|---|---|
| ANGERS | 0.38 | 0.37 | 0.64 | 0.72 | 0.82 |
| NICE | 1.59 | 1.91 | 2.24 | 2.35 | 2.18 |
| TOURS | 0.93 | 0.94 | 0.93 | 0.91 | 0.87 |
| BREST | 0.69 | 0.72 | 0.82 | 1.14 | 1.25 |
| SAINT-ÉTIENNE | 0.37 | 0.50 | 0.48 | 0.85 | 0.87 |
| METZ | 0.20 | 0.24 | 0.54 | 0.61 | 0.56 |
| PAU | 0.16 | 0.22 | 0.35 | 0.40 | 0.39 |
| LE MANS | 0.21 | 0.27 | 0.30 | 0.33 | 0.32 |
| AVIGNON - NIMES | 0.29 | 0.38 | 0.46 | 0.62 | 0.73 |
| PERPIGNAN | 0.26 | 0.28 | 0.30 | 0.33 | 0.27 |
| TOULON | 0.17 | 0.22 | 0.24 | 0.28 | 0.28 |
| CHAMBERY | 0.08 | 0.07 | 0.16 | 0.28 | 0.31 |
| LA ROCHELLE | 0.03 | 0.04 | 0.17 | 0.24 | 0.29 |
| BELFORT | 0.01 | 0.00 | 0.05 | 0.03 | 0.04 |
| TROYES | 0.01 | 0.04 | 0.10 | 0.23 | 0.23 |
| ANNECY | 0.08 | 0.11 | 0.18 | 0.17 | 0.14 |

Source: 1979-89 (OST-Montréal-LISST, 2012), 1999-17 (IRIT-LISST, 2019)

Brest, home to the largest part of the University of Western Brittany (whose other establishments are in Quimper, Morlaix, Rennes, Saint-Brieuc, and Vannes), is also emerging as a relatively important scientific centre. More generally, most of the cities in this category are experiencing an increase in their contributions to French publications.

We do not go into further detail here about the smaller sites[g]. Although underestimated in bibliographic databases because researchers from these sites tend to sign with the address of their laboratory, often based at a university in a larger city, they have also gained in importance over the years [32].

In summary, the analysis of the publications shows a continuous process of deconcentration relative to the Paris region, and more recently, to the major academic centres. This process has slowed down in the 2010's, but it is still ongoing. We hypothesise that it is the result of the increase in the number of university staff, itself linked to the increase in the number of students.

In order to get a clearer picture of the causes of this spatial deconcentration, we now relate publications to the number of researchers and students.

## 3 Just hire more to publish more

By linking scientific production and academic staff, we propose to test the abovementioned theory that productivity and research excellence are the privilege of metropolises with a "critical mass" of researchers sufficient to observe cumulative effects on scientific production.

This theory is historically grounded on the concepts of agglomeration economies and urbanisation economies, which were cast to explain the principles of spatial concentration of certain economic activities [6]. Its transposition to the case of research activities concerns a branch of the regional economy more specifically focused on the study of knowledge spillovers [33,34]. This branch is now more particularly interested in relations and exchanges of measurable knowledge between places [35,36]. However, the idea of a competitive advantage of the largest cities in terms of research has recently come back into force in the academic literature. Researchers specializing in the study of complex systems have transposed a model from biology to highlight an effect of city size on the spatial concentration of research activities, a spatial concentration deemed necessary for scientific performance [37,38].

To relate scientific production to research staff, we use several sources, calculated at the urban agglomeration level: publications in 2016 (in fact a moving average over 2015-17), this time for all disciplines[h]; HER staff at CNRS units in 2018, in the 43 university cities with CNRS units; number of students in 2015-17; HER personnel listed in the "annual social data declarations" (DADS) from 2012 to 2014[i].

---

[f] Ajaccio is the seat of the territorial assembly of Corsica but it does not host a university: the University of Corsica is located in Corte.

[g] A detailed study is under way in Occitania region.

[h] The restriction to SCI Exp of the first part is due to the desire to continue the time series that we had started on this basis. We only have data for all disciplines for the period 1999-2017.

By taking into account all the sites (excluding territories and overseas departments) and several variables simultaneously, the statistical models make it possible to test the link between scientific production and academic staff per site. As the distributions of values are all log-normal, linear regressions can be used, provided we transform the variables with the natural logarithm function. This is a usual procedure in this case to fit to Gaussian or close to Gaussian distributions.

The first model tested covers 114 sites with HER staff and researchers included in the DADS 2014 data called "academic staff" (Fig. 1, Table 6). This model explains about 92% of the variation in the number of publications between urban areas.

**Fig. 1.** Academic staff in 2014 and publications in 2016

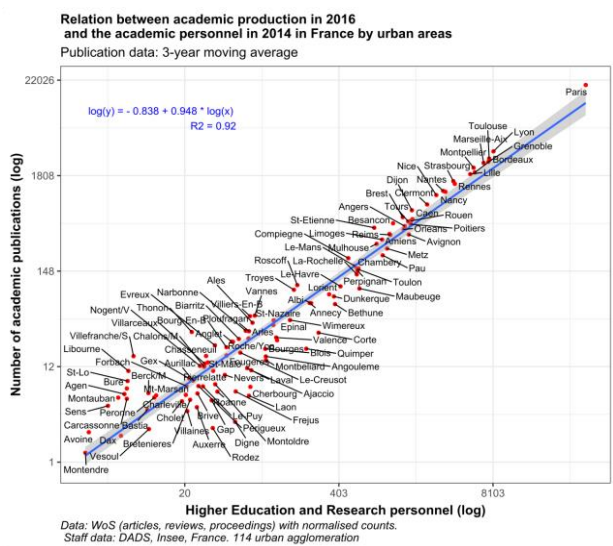

**Table 6.** Summary of regression model: Scientific output in 2016 ~ Academic staff in 2014

| Coefficients | β | Std. error | Statistic t | P. value |
|---|---|---|---|---|
| *(Intercept)* | *-0.838* | *0.141* | *-5.945* | *3.20E-08* |
| log(*Academic staff*) - 2014 | *0.948* | *0.026* | *36.249* | *1.05E-63* |

**Dependent variable**: log(*publis_151617*) 3-y. moving average
**Residual standard error**: 0.5994 on 112 degrees of freedom
**Multiple R-squared**: 0.9215, Adjusted R-squared: 0.9208
**F-statistic**: 1314 on 1 and 112 DF, p-value: < 2.2e-16
**Number of observations**: 114 urban agglomerations
**Sources**: WoS (articles, reviews, proceedings); Insee DADS

---

[i] See the description of private and public jobs and wages in 2015: https://www.insee.fr/fr/statistiques/3536754. We use PCS 342B ("Professors and lecturers"); 342C ("Associate and certified teachers - secondary school teachers - in higher education"); 342D ("Temporary teaching staff in higher education"); 342F ("Directors and research officers in public research"); 342G ("Research and study engineers in public research") and 342H ("Public research fellows") in the "employees" file. Déclaration Annuelle de Données Sociales (DADS): Customized tabulation, INSEE [producer], ADISP [publisher].

No large sites deviate significantly from the model. The few significant deviations concern small sites such as Digne-les-Bains, Villefranche-sur-Saône, and Rodez, for which we have already mentioned that it was complicated to measure scientific output due to the practices of signing articles. It can therefore be concluded that the number of researchers explains almost all of the variations between agglomerations in terms of the number of publications.

Finally, we can go a little further by limiting ourselves to the 43 sites with CNRS employees (Table 7). This time, more information can be taken into account: the number of CNRS researchers, the number of permanent non-CNRS researchers in CNRS units, the number of non-permanent researchers in CNRS units (fixed-term contracts including post-docs), the number of students, and academic staff (moving average on DADS data 2012-2014). The explanation is then 98%.

**Table 7.** Summary of the multiple regression model applied to CNRS sites: scientific production in 2016 ~ researchers and non-permanent staff of CNRS units in 2018, students in 2016, academic staff in 2013

| Coefficients | β | Std.error | Statistic t | P-value |
|---|---|---|---|---|
| (Intercept) | -1.953 | 0.388 | -5.032 | 1.28E-05 *** |
| log(*Academic_staff*) 2012-2014 Insee | 0.508 | 0.129 | 3.936 | 3.52E-04 *** |
| log(*CNRS_res*) 2018 CNRS units | 0.190 | 0.056 | 3.410 | 1.59E-03 ** |
| log(*Non_perm_staff*) 2018 CNRS units | 0.181 | 0.125 | 1.455 | 1.54E-01 |
| log(*Other_res*) 2018 CNRS units | -0.259 | 0.119 | -2.186 | 3.52E-02 * |
| log(*Students*) 2015-2017 MESRI | 0.419 | 0.114 | 3.670 | 7.61E-04 *** |

**Dependent variable**: log(*publis_151617*) 3-y. moving average
**Residual standard error**: 0.2399 on 37 degrees of freedom
**Multiple R-squared**: 0.9793, Adjusted R-squared: 0.9765
**F-statistic**: 350.8 on 5 and 37 DF, p-value: < 2.2e-16
**Number of observations**: 43 urban agglomerations
**Sources**: WoS (articles, reviews, proceedings); Insee DADS 2014-2014; MESRI data on students 2015-2017; data on CNRS research units 2018

No site deviates significantly from this model—leaving aside the case of Corte—but there are some peculiarities that explain its situation (Fig. 2). The statistical model could certainly be improved by including other data (e.g., company research) to get closer to the 100% explanation.

Our results show that the overall number of publications of a site can be accurately predicted from indicators on the number of staff engaged in academically oriented research (corporate researchers publishing significantly less).

In addition, we also used an alternative approach of calculating a publication rate per full-time equivalent researcher (DADS). This additional analysis shows that the large sites all fall within a fairly narrow range, the small sites differ according to their speciality, and an analysis of variance on the types of sites (in four categories) shows a total absence of significance of site size on research productivity.

**Fig. 2.** Standardized residuals from the multiple regression model applied to CNRS sites

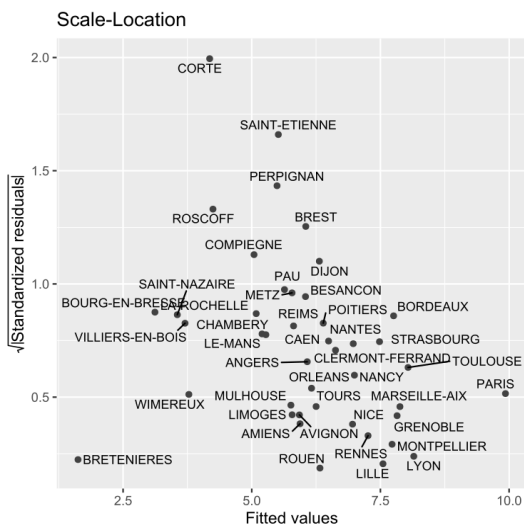

Contrary to many preconceived ideas about the unequal geographical distribution of scientific excellence, the geographical concentration of researchers has no effect on their "productivity" as measured by the number of publications.

## 4 Conclusion

There are limitations to this analysis that should be noted. First, it combines all disciplines, limiting itself to the natural and technical sciences in the first part, and taking into account all academic disciplines in the second part. To go further, however, a per-discipline study is necessary but that requires counting the per-discipline number of researchers.

Second, our study is limited by the source used to estimate scientific production. It is one of the oldest bibliographical databases, whose biases are well documented [39–41] : English-speaking tropism, failure to take into account books, sometimes arbitrary selection of journals taken into account [42]. The human and social sciences in particular would require a specific analysis.

However, despite these limitations, this analysis of the French case is consistent with the main results already observed worldwide: geographical deconcentration and lack of agglomeration effect on the number of publications [24,25,43,44] .

More specifically on the question of the effects of policies implemented since 2007, there is little evidence of a change in the structural trends at work. Whether or not their institutions are grouped together, labelled by the PIA (Plan d'Investissement d'Avenir, that is Future Investment Plan) or not, researchers whatever their city publish with similar intensity.

It may be too early to perceive the effects of policies that have kept researchers busy in recent years and have sought to allocate more research resources to the groups deemed to be the most successful. Much ado about nothing?

Maybe not. Several social movements are underway in the French scientific system challenging the current reforms, which is nothing new or surprising in itself, whether or not these reforms or challenges are considered well-founded [45]. All these movements report the waste of energy and time devoted to the reorganisation of their institutions, in particular all the mergers of establishments that have been underway for more than ten years (one needs only to follow the discussions on the changing contours of the Communities of Establishments in the Paris region or look at the work that analyses the complex interplay of actors involved in these transformations [46,47]). Above all, these reconfigurations would have led to a multiplication of organizational levels and a loss of legibility of the French higher education and research system, as shown by "the layer cake of the organizational affiliations" [48,49], a paradoxical result given the initially stated objective: to gain international visibility and move up in international rankings. If we consider, as we have argued in this text, that the analytical bases that underpin part of the public justification of the policies in question are erroneous and contradict long-term trends that predominate internationally, then it is possible to interpret this dissonance by invoking the notions of complexity and disorder.

That the French scientific system is complex, among the most complex in the world because of its history, is beyond doubt for all those who know it. This system evolves according to trends well identified in our work and in many other similar works, among which is the spatial deconcentration of activity. This can be seen as a form of order. Disrupting an evolving complex system is always possible. Public policy can consciously seek to do so, for example, by putting in place mechanisms to counter the trend of increasing inequality in a school system. The interesting feature of the HER policies we have discussed in this paper is that they are based on misinterpretations of trends: believing in an inevitable movement towards spatial concentration of innovation and research activities in the most populous cities whilst believing in greater scientific productivity in these cities. Their designers believe that they can accompany and reinforce these trends and use them to enhance the efficiency of the system, whereas their policies appear to be completely out of step with the observation of trends provided by empirical analyses such as ours. Their interventions then appear as a kind of more or less random disturbance: a disorder. For the time being, we can simply say that they introduce tensions into the system. It is still too early to know if they will change the trajectory of the system and in which direction.

**Acknowledgements:** We thank PROGEDO-ADISP for sharing social data on HER staff. The scripts and the data (except for the social data concerned by statistical secrecy) are released at: https://framagit.org/MarionMai/the-spatial-organisation-of-french-research